\documentclass[aps,preprint]{revtex4}
\usepackage{pifont}
\usepackage{mathrsfs}
\usepackage{amsfonts}
\usepackage{amsmath}
\usepackage{amssymb}
\usepackage{graphicx}%
\begin{document}
\title{From quantum to classical description of intense laser-atom physics
with Bohmian trajectories}
\author{X. Y. Lai$^{1}$}
\author{Qingyu Cai$^{1}$}
\email{qycai@wipm.ac.cn}
\author{M. S. Zhan$^{1,2}$}
\affiliation{$^{1}$State Key Laboratory of Magnetic Resonances and
Atomic and Molecular Physics, Wuhan Institute of Physics and
Mathematics, The Chinese Academy of Sciences, Wuhan 430071, China}
\affiliation{$^{2}$Center for Cold Atom Physics, The Chinese Academy
of Sciences, Wuhan 430071, China}

\begin{abstract}
In this paper, Bohmian mechanics is introduced to the intense
laser-atom physics. The motion of atomic electron in intense laser
field is obtained from the Bohm-Newton equation. We find the quantum
potential that dominates the quantum effect of a physical system
becomes negligible as the electron is driven far away from the
parent ion by the intense laser field, i.e. the behavior of the
electron smoothly trends to be classical soon after the electron was
ionized. Our numerical calculations present a direct positive
evidence for the semiclassical trajectory methods in the intense
laser-atom physics where the motion of the ionized electron is
treated by the classical mechanics, while quantum mechanics is
needed before the ionization.
\end{abstract}

\pacs{PACC numbers: 0365, 3280K}

\maketitle

\section{Introduction}

In recent years, intense laser-atom physics has received more and
more attention \cite{bk}, due to the nonlinear multiphoton
phenomena, such as high-order harmonic generation and
above-threshold ionization. Without doubt, such multiphoton
phenomena could be reproduced by resolving the time-dependent
Schr\"{o}dinger equation (TDSE). In order to understand such
phenomena intuitively, some semiclassical approaches, mixing
classical and quantum arguments, have been proposed
\cite{Corkum,Corkum2,Schafer,Paulus}. In the semiclassical
approaches, the motion of ionized electrons is treated by classical
dynamics directly, while quantum mechanics is used before the
ionization. These semiclassical approaches have obtained much
success in intense laser-atom physics, although there is no explicit
evidence for their key assumption that the ionized electron can be
treated by the classical mechanics directly \cite{lm}.

Bohmian mechanics (BM) \cite{bd,hpr,nh} also called quantum
trajectory method \cite{rew} is another version of quantum theory,
in which trajectory concept is used to describe the motion of
particles with the Bohm-Newton equation. It has been successfully
used to study some fundamental quantum phenomena such as the
two-slit experiment \cite{cp} and tunneling \cite{joh,cd}. Recently,
Oriol \emph{et al} \cite{xo} used BM to simulate the electron
transport in mesoscopic systems. Sanz \emph{et al} \cite{ass1}
applied this theory to the atom surface physics. Makowski \emph{et
al} derived some central potentials \cite{maj1} and two-dimensional
noncentral potentials \cite{maj2} from BM to investigate the exact
classical limit of quantum mechanics \cite{rosen}. BM has also been
regarded as a resultful approach to studying chaos \cite{us,ce}, due
to its description of trajectory for the quantum system. It has also
been successfully applied to the intense laser-atom physics to study
the dynamics of above-threshold ionization \cite{lxy1} and
high-order harmonic generation \cite{lxy2} by authors.

The only difference between the Bohm-Newton equation and Newton
equation is that there is an extra term in the Bohm-Newton equation,
called quantum potential. When quantum potential is negligible, the
Bohm-Newton equation will reduce to the standard Newton equation and
then the motion of particles can be described by classical
mechanics.

In this paper, BM is introduced to the intense laser-atom physics.
We first obtain the electron quantum trajectories of an atomic
ensemble from the Bohm-Newton equation. Next we study how the
quantum potentials change as the electrons are driven away from the
parent ion by the laser field. We find that quantum potentials trend
to be smaller and then become completely negligible soon after the
electrons were ionized. In this case, the Bohm-Newton equation
reduces to Newton equation, and then the motion of electrons can be
described by classical equation of motion. Thus, our results present
a direct positive evidence for the previous semiclassical trajectory
methods in intense laser-atom physics where the motion of ionized
electrons is treated by classical mechanics directly,  while quantum
mechanics is needed before the ionization. On the other hand, our
numerical results clearly show how far the electron from the core as
it was ionized, while the initial position of the ionized electron
is usually chosen to be 0 (the position of the core) in the
semiclassical approaches \cite{Corkum}.

\section{Quantum trajectories formalism
of Bohmian mechanics}\label{secbm}

BM \cite{bd,hpr} is derived from a subtle transformation of the
time-dependent Schr\"{o}dinger equation. Firstly, a wavefunction can
be
written in the polar form $\psi(\mathbf{r},t)=R(\mathbf{r},t)e^{iS(\mathbf{r}%
,t)/\hbar}$, where $R$ and $S\ $are real functions. Secondly, the
wavefunction is inserted into the time-dependent Schr\"{o}dinger
equation.
The real part of the resulting equation has the form%
\begin{equation}
\frac{\partial S}{\partial t}+\frac{\left( \nabla S\right) ^{2}}{2m}+V-\frac{%
\hbar^{2}}{2m}\frac{\nabla^{2}R}{R}=0   \label{qp}
\end{equation}
and the imaginary part is
\begin{equation}
\frac{\partial\rho}{\partial t}+\nabla(\rho v)=0,   \label{qc}
\end{equation}
where $\rho(\mathbf{r},t)=R^{2}(\mathbf{r},t)$, $v=\nabla S(\mathbf{r}%
,t)/m$, and $V$ is the ordinary potential. Equation (\ref{qp}) is
similar to the classical Hamilton-Jacobi
equation, except it has an extra term, $-\frac{\hbar^{2}}{2m}\frac{%
\nabla^{2}R}{R}$. This term, denoted by $Q(\mathbf{r},t)$ in this
paper, is usually called the quantum potential. Equation (\ref{qc})
looks like the classical continuity equation. So a Bohm-Newton
equation of motion for a Bohmian particle can be constructed:
\begin{equation}
md^{2}\mathbf{r}/dt^{2}=-\nabla(V+Q)   \label{bn}
\end{equation}
from the standpoint of classical mechanics. The motion of particle
is determined by the ordinary potential $V$ and the quantum
potential $Q$ which plays a crucial role for the appearance of
quantum phenomena. Obviously, when quantum potential $Q\rightarrow
0$ in equation (\ref{bn}), the Bohm-Newton equation will reduce to
the standard Newton equation and then the motion of particle can be
described by classical mechanics.

In fact, the trajectory of particles can be evolved from a much
simpler equation of motion instead of equation (\ref{bn})
\cite{bd,hpr}:
\begin{equation}\label{bm}
\frac{d\mathbf{r}}{dt}=\nabla S(\mathbf{r},t)/m.
\end{equation}
To obtain the trajectory, we need to know the phase
$S(\mathbf{r},t)$ and the initial positions of particles. According
to BM, the initial distribution of particles in an ensemble is given
by $|\psi(\mathbf{r},0)|^{2}$.

\section{Intense laser-atom interaction}

\begin{figure}
  \includegraphics[width=5in]{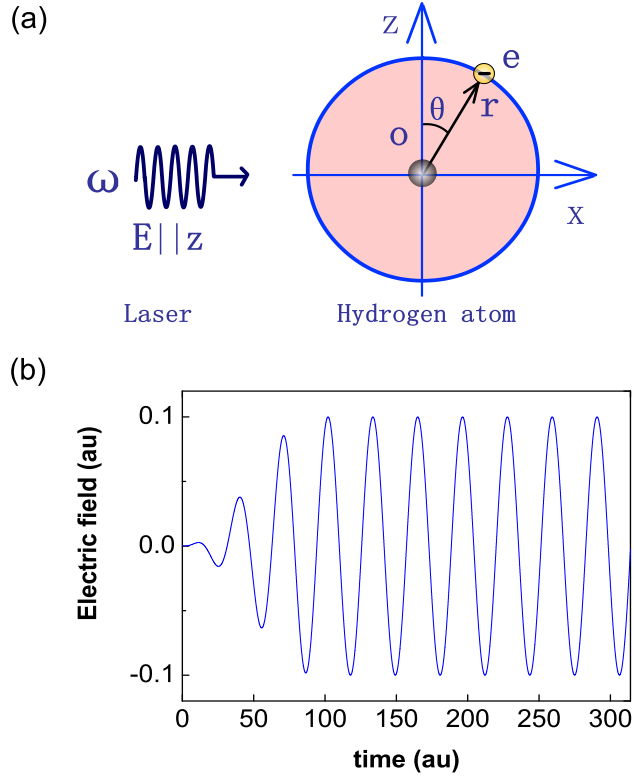}\\
  \caption{ (color online). (a) Graphic description of the direction
of the linearly polarized laser field ($\mathbf{E}||\mathbf{z}$) at
frequency $\omega$ and the coordinate of the Hydrogen system; (b)
Laser field profile in the first ten periods, where  the laser
period $T=10\pi$ in our work. }\label{1}
\end{figure}

The system we consider in this paper is a Hydrogen atom in intense
laser field. The Schr\"{o}dinger equation for the system can be
written as (atomic units are used throughout)
\begin{equation}\label{tdse}
i\frac{\partial\psi(\mathbf{r},t)}{\partial t}=[H_{0}(%
\mathbf{r})+H_{I}(\mathbf{r},t)]\psi(\mathbf{r},t).
\end{equation}
Here $H_{0}(\mathbf{r})$ is the field-free Hydrogen atom Hamiltonian
and $H_{I}(\mathbf{r},t)$ is the
intense laser-atom interaction: $H_{0}(\mathbf{r})=-\frac{1}{2}\frac{d^{2}}{%
dr^{2}}+\frac{\widehat{L}^{2}}{2r^{2}}-\frac{1}{r},H_{I}(\mathbf{r},t)=-\mathbf{%
r\cdot E}(t)=-zE(t)$, where the laser is the linearly polarized field ($%
\mathbf{E}||\mathbf{z}$) and $E(t)$ is the laser field profile. Due
to the linearly polarized laser field, magnetic quantum number $m$
of the atom is a good
quantum number, so that the problem of solving the time-dependent Schr\"{o}%
dinger equation here can be simplified into a two-dimensional
problem (see figure \ref{1}(a)). In our study we use the grid method
and the second-order split-operator technique to numerically solve
the time-dependent Schr\"{o}dinger equation (\ref{tdse}), which has
been detailedly introduced by Tong and Chu \cite{tong}. The laser
field profile is
\begin{equation}
E(t)= \left\{
\begin{array}{ll}
E_{0}\sin^{2}(\frac{\pi t}{6T})\sin(\omega t), & {0\leq t\leq 3T} \\
E_{0}\sin(\omega t), & {\ t> 3T,}%
\end{array}
\right.
\end{equation}
 where $T=2\pi/\omega$, $E_{0}$ and $\omega$ are the
electric field amplitude and angular frequency, respectively (see figure \ref{1}(b)). Here we take $%
E_{0}=0.1$ au and $\omega=0.2$ au The initial state
$\psi(\mathbf{r},0)$ of the system is in the ground state of the
field-free Hydrogen atom.

\section{Result}

\subsection{Electron trajectory from the Bohm-Newton equation}

\begin{figure}
  \includegraphics[width=5in]{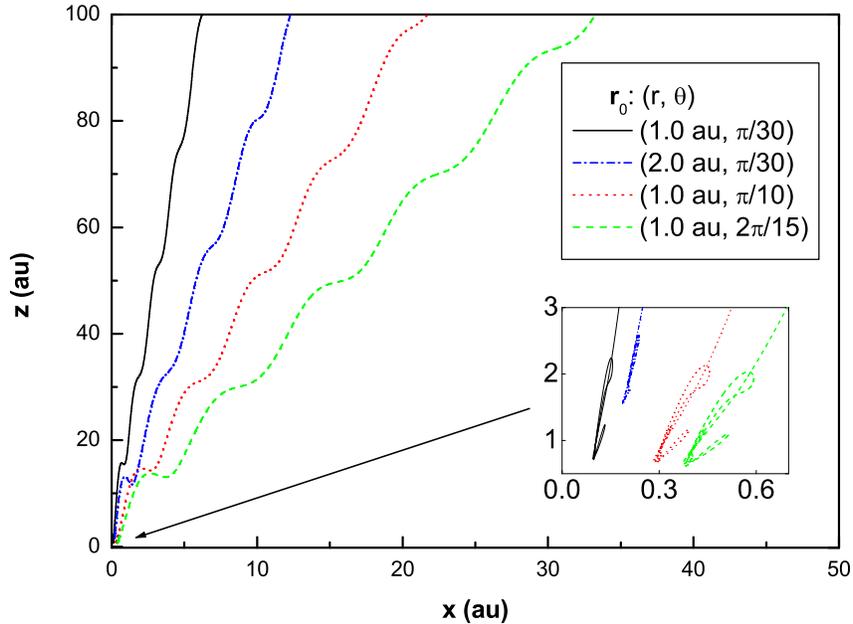}\\
  \caption{(color online). The electron trajectories in the spatial
coordinate with the corresponding initial positions
$\mathbf{r}_{0}$. The initial positions $\mathbf{r}_{0}$ in the
polar coordinate are from top to bottom: $(1.0 $ au$, \pi/30)$,
$(2.0 $ au$, \pi/30)$, $(1.0 $ au$, \pi/10)$, and $(1.0 $ au$,
2\pi/15)$, respectively.}\label{2}
\end{figure}

After numerically obtaining the time-dependent wavefunction
$\psi(\mathbf{r},t)$, and then the phase $S(\mathbf{r},t)$, we can
get the trajectory of an electron
by integrating equation (\ref{bm}) with its initial value of position $\mathbf{r}%
_{0}$ (see appendix A). In this way, we can gain an ensemble of
electron trajectories according to the initial distribution of
electrons $\left\vert \psi(\mathbf{r},0)\right\vert ^{2}$, where
$\psi(\mathbf{r},0)$ is the ground state of the field-free Hydrogen
atom. Explicitly, we present four trajectories of electron with the
corresponding initial positions in figure \ref{2}. Let's take one
electron trajectory as an example with $\mathbf{r}_{0}=(1.0$ au,
$\pi/30)$ in the polar coordinate. The path of the electron has the
following character in the spatial coordinate: The motion is
irregular when the electron is near the parent ion; after the
electron has been driven far away from the parent ion by the intense
laser field, it travels in a straight line with a little
oscillation. The reason why the motion of the electron keeps
oscillating as the electron is far away from the parent ion is due
to the existence of the laser field. The other trajectories with
different initial positions have similar character as shown in
figure \ref{2}.

\begin{figure}
  \includegraphics[width=5in]{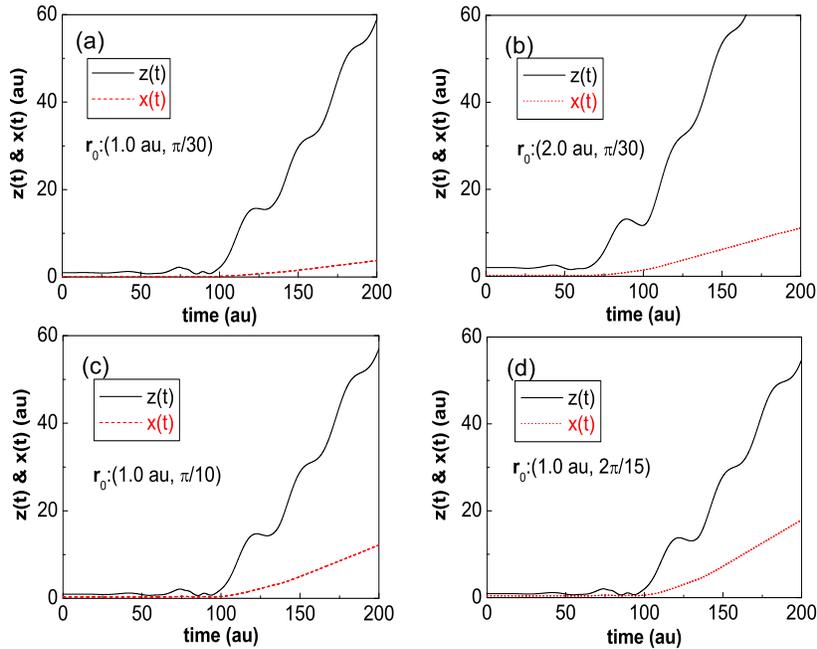}\\
  \caption{(color online). Projections of electron trajectories in $z$ and $%
x$ directions as functions of time. The corresponding initial
positions of electrons are, from (a) to (d),   $(1.0 $ au$,
\pi/30)$, $(2.0 $ au$, \pi/30)$, $(1.0 $ au$, \pi/10)$, and $(1.0 $
au$, 2\pi/15)$, respectively. It is obvious that the motion of the
electron in $x$ direction is uniform while its motion in $z$
direction has undulations as the electron is far from the core. The
reason for the $z$-undulation is due to the existence of laser field
($\mathbf{E||z}$). }\label{3}
\end{figure}

In figure \ref{3}, we show the projections of electron trajectory in $z$ and $%
x$ directions, respectively, as functions of time. In figure
\ref{3}(a), two curves $z(t)$ and $x(t)$
 are obtained from the electron trajectory with $\mathbf{r}_{0}=(1.0$ au,
 $\pi/30)$. At the beginning, both $z(t)$ and $x(t)$ are near the core.
After the time $t\approx130$ au, $z(t)$ becomes a regular wave curve
and $x(t)$ seems to be a straight line. Similarly, we have obtained
the projections of three other electron trajectories with their
initial positions $\mathbf{r}_{0}$ (see figures 3(b)-(d)). All of
them have the similar character as that in figure \ref{3}(a).

\subsection{Time-dependent quantum potential in intense laser-atom
physics}

\begin{figure}
  \includegraphics[width=5in]{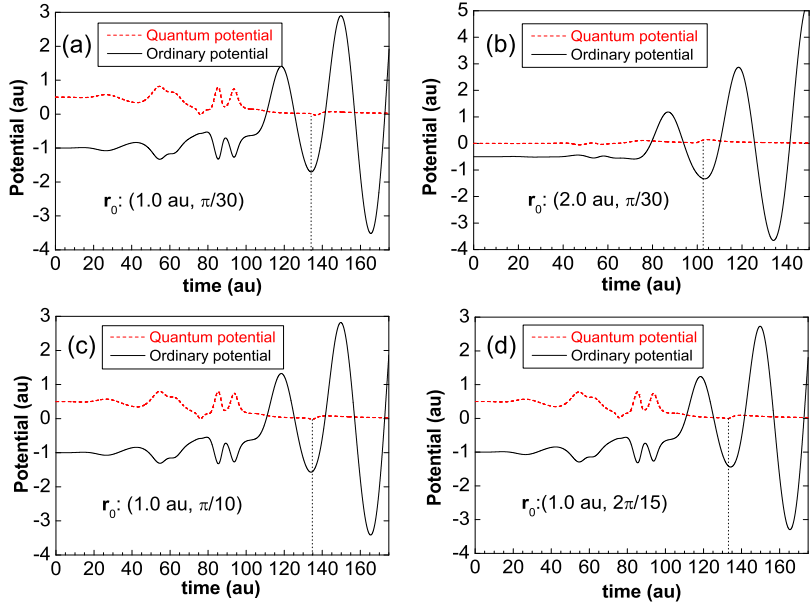}\\
  \caption{(color online). The time-dependent value of quantum potential
   and ordinary potential of the electrons with the corresponding initial positions,
    $(1.0 $ au$, \pi/30)$, $(2.0 $
au$, \pi/30)$, $(1.0 $ au$, \pi/10)$, and $(1.0 $ au$, 2\pi/15)$,
respectively.}\label{4}
\end{figure}

In Section \ref{secbm}, we have discussed that quantum potential
$Q(\mathbf{r},t)$ together with ordinary potential $V(\mathbf{r},t)$
dominates the motion of particle in BM. And quantum potential plays
a crucial role for the appearance of quantum phenomenon. If the
value of $Q(\mathbf{r},t)$\ smoothly trends to be negligible
($Q\rightarrow 0$), equation (\ref{bn}) will reduce to Newton
equation and then the motion of electron can be described by
classical mechanics.

Now we study the change of quantum potential for an atomic electron
as it is driven away by the laser field. After obtaining the
time-dependent wavefunction $\psi(\mathbf{r},t)$ and the electron
trajectories $\mathbf{r}(t)$, we calculate the changes of quantum
potential $Q(\mathbf{r}(t))$ (see appendix B). In figure \ref{4}, we
show the time-dependent values of quantum potential
$Q(\mathbf{r}(t))$ and ordinary potential
$V(\mathbf{r}(t))=-\frac{1}{r}-zE(t)$ with the corresponding initial
positions $\mathbf{r}_{0}$. Here we take the electron with
$\mathbf{r}_{0}=(1.0$ au, $\pi/30)$ as an example (see figure 4(a)).
The value of quantum potential is comparable with that of ordinary
potential in the early period. After some time, \emph{e.g.}, after
100 au about, quantum potential smoothly becomes smaller and then it
trends to be negligible, particularly after the time $t\approx 133$
au The changes of quantum potential of three other electrons are
presented in figures 4(b)-(d), and they have the similar character
as that in figure \ref{4}(a). In this way, our numerical results
present an explicit picture of the reduction from the Bohm-Newton
equation to Newton equation. Consequently, the Bohmian trajectory
method can give both the quantum and classical descriptions of the
motion of an atomic electron in intense laser-atom physics.

\subsection{Classical description after the electron has been ionized}

\begin{figure}
  \includegraphics[width=5in]{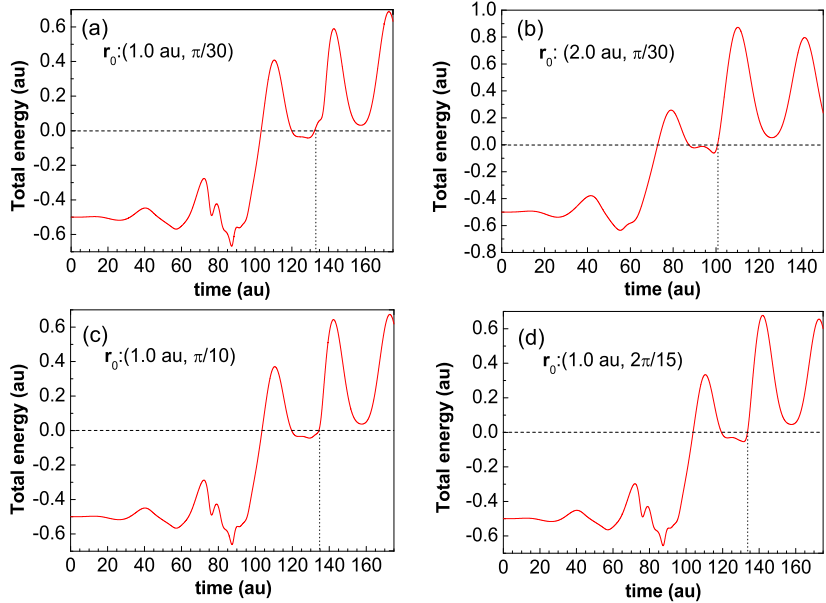}\\
  \caption{The time-dependent total energies of an electron (including kinetic energy,
   Coulomb potential and quantum
potential) with the initial positions,
  $(1.0 $ au$, \pi/30)$, $(2.0 $
au$, \pi/30)$, $(1.0 $ au$, \pi/10)$, and $(1.0 $ au$, 2\pi/15)$,
respectively.}\label{5}
\end{figure}

As we have shown above, the Bohm-Newton equation of electron is
approaching to Newton equation in intense laser-atom physics, as the
electron was driven far away from the parent ion. In the previous
semiclassical approaches of intense laser-atom physics, the ionized
electron motion is usually treated by the classical dynamics, while
quantum mechanics is used before the ionization
\cite{Corkum,Corkum2,Schafer,Paulus,Salieres}. Here, we want to
verify whether the motion of an electron can be treated by classical
dynamics or not, i.e. whether the quantum potential becomes
negligible or not, as it has been ionized.

In figure \ref{5}, we show the total energies of an electron
(including kinetic energy, Coulomb potential and quantum potential)
with different initial positions. When the total energy becomes
positive, the electron will be ionized. Here we take the electron
with $\mathbf{r}_{0}=(1.0$ au, $\pi/30)$ as an example (see figure
5(a)). At the beginning, the value of the total energy is less than
zero, i.e. the electron is bounded by the core. After the time
$t\approx103$ au, the total energy becomes positive but it then
turns to the negative at the time $t\approx120$ au The reason why
the total energy of the ionized electron turns to the negative again
may be that the electron with low energy is not far enough from the
core and it is recaptured by the parent ion with a photon emitted
when the laser field reverses.
The total energy becomes positive again and higher than before at
$t\approx133$ au and thereafter, i.e. the electron can never be
recaptured again by the parent ion. On the other hand, in figure
\ref{4}(a), we found the value of quantum potential of the electron
trends to be trivial soon after the time $t\approx133$ au (The
corresponding distance between the ionized electron and the core is
16.2 au about; see figure 3). Similarly, in figures 5(b)-(d) and
4(b)-(d), we can find when the electron is ionized, the
corresponding quantum potential trends to be negligible. Therefore,
our numerical solutions show that the quantum potential is
negligible as the ionized electron was driven far away from the core
and thus the motion of ionized electron is dominated by classical
mechanics. In this way, our work may present a direct positive
evidence for the semi-empirical assumption in the semiclassical
approaches of intense laser-atom physics, such as the three-step
model \cite{Corkum} and Feynman's path-integral approach in the
strong field approximation \cite{Salieres}, in which the motion of
ionized electrons is treated by classical dynamics directly.
Furthermore, the Bohmian trajectory method can help us to find how
far the electron from the core when it was ionized, while the
initial position of an ionized electron is assumed to be 0 in the
semiclassical approaches \cite{Corkum}.

\section{Discussion}

As it has been discussed above, quantum potential dominates the
quantum behavior of a physical system. In the prior work of
quasistatic model of strong-field multiphoton ionization
\cite{Corkum}, a dual procedure was considered to describe the
multiphoton phenomena in intense laser-field physics. First a
tunneling model is applied to obtain the ionization rate and
describe the formation of a sequence of wave packets. The second
part of the quasistatic procedure uses the classical mechanics to
describe the evolution of an electron wave packet. It has been
pointed out definitely that only the Newton equation cannot give the
ionization rate correctly, unless a quantum tunneling process was
allowed \cite{jsc}. This is consistent with our numerical results in
this paper: The quantum potential cannot be ignored before the
ionization, i.e. quantum mechanics is needed to study the behavior
of the electron before the ionization, but the motion of an ionized
electron can be described by classical mechanics directly .

\section{Remark}

It is interesting that Mahmoudi \emph{et al} \cite{mah} have studied
free-electron quantum signatures in intense laser fields by looking
for negativities in the Winger distribution. They concluded that the
quantum signatures of the free electron get washed away, provided
the dipole approximation is made. In this way, their work presents
another evidence for the treatment of ionized electron in the
semiclassical approaches of intense laser-atom physics.

\section{ Acknowledgement}

We thank J. B. Delos and L. You for helpful discussions. This work
is supported by National Basic Research Program of China under Grant
No. 2006CB921203.

\appendix

\section{Electron quantum trajectory}

In this appendix, we show how to obtain the electron quantum
trajectory from equation (\ref{bm}). First, equation (\ref{bm}) can
be expressed as (atomic units are used)
\begin{equation}\label{a1}
\frac{d\mathbf{r}}{dt}=\nabla S(\mathbf{r},t)=\text{Im}
\frac{\nabla\psi(\mathbf{r},t)}{\psi(\mathbf{r},t)},
\end{equation}
where $\psi(\mathbf{r},t)=R(\mathbf{r},t)e^{iS(\mathbf{r}%
,t)}$ and $\text{Im}(f)$ is the imaginary part of $f$. In our study
$\psi(\mathbf{r},t)$ is gained by  numerically solving the
time-dependent Schr\"{o}dinger equation (\ref{tdse}), using the grid
method and the second-order split-operator technique, which has been
detailedly introduced by Tong and Chu \cite{tong}. Then we use the
Runge-Kutta method to evolve equation (\ref{a1}) to obtain the
electron quantum trajectory.

In our numerical procedure, the range of the variable $\mathbf{r}$
in the radial direction  is confined to $(0,150$ au$)$. In the grid
method, the numbers of grid points are $350$ in the $r$ direction
and $51$ in the $\theta$ direction. The time step for the evolution
is $0.01$ au

\section{Time-dependent value of quantum potential}
After obtaining the time-dependent wavefunction $\psi(\mathbf{r},t)$
and the electron quantum trajectory $\mathbf{r}(t)$ with the
corresponding initial position $\mathbf{r}_{0}$, we can numerically
obtain the value of the quantum potential $Q(\mathbf{r}(t))$ along
the trajectory $\mathbf{r}(t)$. First, the quantum potential
$Q(\mathbf{r},t)$ can be written as
\begin{equation}\label{b1}
Q(\mathbf{r},t)=-\frac{1}{2}\frac{%
\nabla^{2}R(\mathbf{r},t)}{R(\mathbf{r},t)}=-\frac{1}{2}\left[
\text{Re}\frac{\nabla^{2}\psi(\mathbf{r},t)}{\psi(\mathbf{r},t)}+\left(\text{Im}
\frac{\nabla\psi(\mathbf{r},t)}{\psi(\mathbf{r},t)}\right)^{2}\right],
\end{equation}
where $\text{Re}(f)$ is the real part of $f$. Secondly, inserting
the quantum trajectory $\mathbf{r}(t)$ into equation (\ref{b1}), we
can numerically get the value of quantum potential
$Q(\mathbf{r}(t))$.

\end{document}